\documentclass[prf,preprint,nobibnotes,longbibliography]{revtex4-1}
\usepackage{epsfig}
\usepackage{amsmath}
\usepackage{fancybox,amssymb}
\usepackage[letterpaper, portrait, margin=1in]{geometry}
\renewcommand{\eqref}[1]{(\ref{#1})}

\begin{document}
\title{\bf Super-Diffusive Gas Recovery from Nanopores\footnote{Article accepted by Phys. Rev. Fluids}}
\author{Haiyi Wu}
\author{Yadong He}
\author{Rui Qiao}
 \email[Email: ] {ruiqiao@vt.edu}
 \homepage{http://www.me.vt.edu/ruiqiao}
 \affiliation{Department of Mechanical Engineering,
   Virginia Tech, Blacksburg, Virginia, 24061, USA}
\date{\today}
\begin{abstract}

Understanding the recovery of gas from reservoirs featuring pervasive nanopores is essential for effective shale gas extraction. Classical theories cannot accurately predict such gas recovery and many experimental observations are not well understood. Here we report molecular simulations of the recovery of gas from single nanopores, explicitly taking into account molecular gas-wall interactions. We show that, in very narrow pores, the strong gas-wall interactions are essential in determining the gas recovery behavior both quantitatively and qualitatively. These interactions cause the total diffusion coefficients of the gas molecules in nanopores to be smaller than those predicted by kinetic theories, hence slowing down the rate of gas recovery. These interactions also lead to significant adsorption of gas molecules on the pore walls. Because of the desorption of these gas molecules during gas recovery, the gas recovery from the nanopore does not exhibit the usual diffusive scaling law (i.e., the accumulative recovery scales as $R \sim t^{1/2}$ but follows a super-diffusive scaling law $R \sim t^n$ ($n>0.5$), which is similar to that observed in some field experiments. For the system studied here, the super-diffusive gas recovery scaling law can be captured well by continuum models in which the gas adsorption and desorption from pore walls are taken into account using the Langmuir model.

\end{abstract}
\maketitle

\section{Introduction}
\label{sec:intro}

Natural gas production from shale formation has received significant attention recently and can potentially lead to a new global energy source. For example, according to the U.S. Energy Administration, shale gas provides the largest source of the growth in U.S. natural gas supply during the past decade, and its share is expected to grow continually in the future \cite{ref1}. To enable the effective extraction of shale gas, accurate prediction of gas recovery from shale formation is needed.

Predicting the gas recovery from shale formations requires a fundamental understanding of the gas transport in these formations. A unique aspect of shale formations is that gas are mostly trapped in ultra-tight rock pores with permeability on the order of nanodarcies \cite{ref2,ref3}. The pore size of these shale matrices spans from a few to several hundred nanometers \cite{ref4,ref5,ref6,ref7,ref8,ref9}, and the number of nanopores in shale gas reservoirs is much larger than that in conventional gas reservoirs \cite{ref10}. Because the size of nanopores can become comparable and even smaller than the intrinsic length scale of the gas (i.e., the mean free path) or the length scale of the molecular interactions between gas molecules and pore walls, gas transport in nanopores can deviate from that in macropores. Much insight has been gained on such transport thanks for the extensive research in the past century \cite{ref11,ref12,ref13,ref14,ref15}. For example, the nature of gas transport in nanopores depends on the Knudsen number $Kn=\lambda/d$ , which is the ratio of the local gas mean free path ($\lambda$) to the pore width ($d$) \cite{ref16,ref17,ref18}. The gas flow is classified into different regimes including continuum flows ($Kn<0.01$), slip flows ($0.01<Kn<0.1$), transition flows ($0.1<Kn<10$); and molecular flows ($Kn>10$) \cite{ref17,ref19}. Except in the continuum flow regime, the no-slip boundary condition and/or the Navier-Stokes equations are not strictly valid. Under these conditions, more advanced models, most of which based on kinetic theories, have been be used. For example, various slip models have been used to describe the slippage effect on the pore walls \cite{ref20,ref21,ref22,ref23}. In addition to the non-classical flow behaviors, prior research also indicated that the Knudsen diffusion can dominate the transport process in nanopores when the Knudsen number is large because the collisions between gas molecules and pore wall are more frequent than collisions between gas molecules \cite{ref24,ref25,ref26,ref27,ref28}. These insights are highly relevant to the gas transport in shale gas formations. For instance, Chen and co-workers constructed the structure of shale formation using the Markov chain Monte Carlo method and simulated the gas transport in the constructed structures using the Lattice Boltzmann method \cite{ref24}. Their simulations indicated that gas flow in shale formation differs greatly from the Darcy flow, and the Knudsen diffusion always plays a role in shale gas transport. Finally, when the pore width approaches the length scale of molecular gas-pore wall interactions (usually on the order of a few nanometers), the gas transport within the pore is strongly affected by these interactions \cite{ref2,ref24,ref26,ref29}.

Building on the fundamental understanding of gas transport in nanopores, many theoretical and simulation studies on the recovery of gas from shale formations have been carried out \cite{ref2,ref30, ref31}. Much of these works showed that the gas recovery from shale formation is effectively a diffusive process: during the early stage of gas recovery, the cumulative gas production obeys a simple scaling law 
\begin{equation}
\begin{array}{r@{}l}
    FR(t) \approx & \ \alpha t^{1/2} \\ 
    PR(t) \approx & \ \beta t^{-1/2} 
\end{array}
\label{eq1}
\end{equation}
where $RF(t)$ is the recovery fraction of the total trapped gas inside the shale formation recovered between time $0$ and $t$, and $PR(t)$ is the gas production rate. $\alpha$ and $\beta$ are pre-factors that depend on the size of pore, the initial pressure of gas inside the pore, etc. During the late stage of gas recovery, the gas production rate decreases exponentially and is often not useful for practical gas recovery operations. While the scaling laws in Equation \eqref{eq1} have indeed been observed in some field experiments, other behavior has also been reported. For example, data compiled from the gas production from 25 wells of Barnett shale\cite{ref32} showed that, during the early stage of gas production, the decay of the production rate approximately follows a power law with an index $-0.4$, the accumulative gas recovery thus observes a super-diffusive scaling law $RF(t) \approx \alpha t^m$ with $m \approx 0.6$. At present, the physical origin of this super-diffusive scaling is not yet well understood. Nevertheless, some studies have shown that the desorption of gas from pore walls tends to make the decay of gas production rate slower, but its effect on the scaling of gas production was not discussed \cite{ref33}.

In this work, we study the gas recovery from single nanopores using molecular dynamic (MD) simulation. In particular, we focus on the scaling laws for gas recovery fraction and the impact of gas-pore wall interactions on gas transport inside the pore and the gas recovery from the pore. Our work is inspired in part by prior studies on shale gas recovery \cite{ref2,ref11,ref31}. While these studies have provided powerful insights into the dynamics of shale gas recovery, some important issues were not yet addressed. For example, since they are based on classical kinetic theories, these studies do not clarify the effects of molecular gas-wall interactions. These interactions are expected to be important in pores narrower than a few nanometers, which are abundant in some shale formations \cite{ref10}. By using MD simulations to study the gas recovery from nanopores, we will explicitly address these issues.

The rest of the paper is organized as follows: In Sec. \ref{sec:model}, we introduce the MD model for gas recovery from nanopores, the MD simulation method, and our simulation protocol. In Sec \ref{sec:results}, we present our simulation results on gas recovery, focusing on its qualitative nature (i.e., scaling law of gas recovery) and its quantitative aspect (i.e., rate of gas production and the total diffusion coefficients of gas inside pores). Finally the conclusions are drawn in Sec. \ref{sec:conclusions}.

\section{MD simulation details}
\label{sec:model}

Simulation of gas recovery from shale formations appears extremely difficult due to the vastly different scales of pore sizes in shale formations (from a few nanometers to millimeters), the diverse structures and surface chemistry of these pores, and the complicated connectivity between these pores. Nevertheless, it has been recognized that the overall gas recovery from the formation is limited by the transport of gas from the narrow pores to large fractures \cite{ref2}. In addition, prior studies have shown that the essential features of gas recovery is captured well by pore scale modeling utilizing simple pore geometries, e.g. cylindrical or slit pores \cite{ref2}. In this work, we study the gas recovery from slit-shaped nanopore with a constant width using MD simulations. While this nanopore model does not take into account the rich structure and chemical properties of pores in shale formations, it does allow us to focus on the generic features of gas recovery in nanopores in shale formations. The insight gained in this study can form a foundation for understanding gas recovery from more realistic pores.

\begin{figure}[hptb]
  \begin{center}
    \epsfig{file=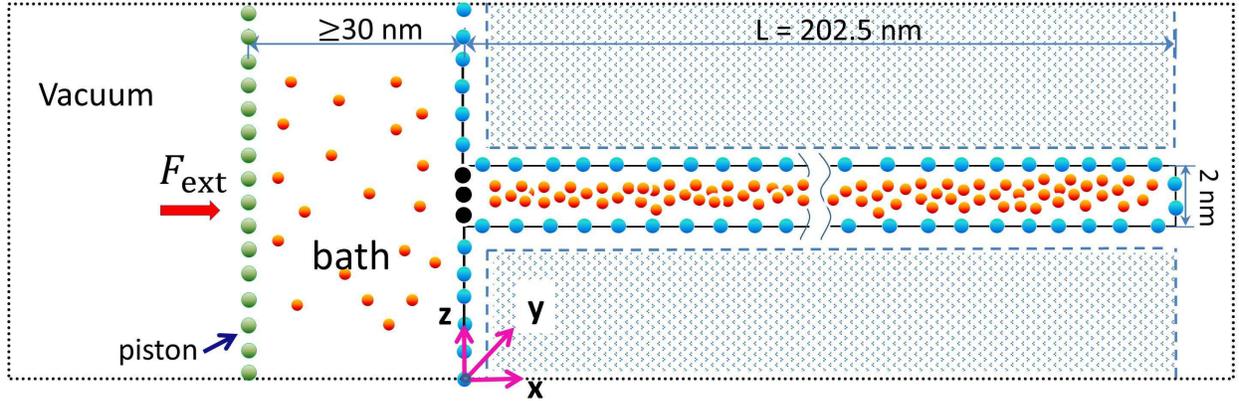,width=6.5in}  
    \caption{A schematic of the nanopore-piston MD system used for studying gas recovery from nanopores. The black dotted line denotes the simulation box, which is periodical in y- and z-directions. The shaded regions denote the implicit pore walls (see text for details). The black spheres at the pore entrance are â€œblockerâ€ atoms that are removed during gas recovery simulations. The green spheres at the left end side are piston atoms.
    }
    \label{fig1}
  \end{center}
\end{figure}

Figure \ref{fig1} shows a schematic of the MD system. The system features a slit pore, a gas bath, a piston, and the methane gas molecules inside the pore and the gas bath. The right end of the pore is permanently sealed. Initially, the left end of the pore is sealed by fixing some â€œblockerâ€ atoms at its outlet (the black spheres in Fig. \ref{fig1}), and gas inside the pore has higher pressure than in the gas bath. During gas recovery simulations, the left end of the pore is made open by removing those â€œblockerâ€ atoms. In this setup, the slit pore can be considered as one of the many pores within a shale-rock matrix, while the gas bath can be considered as the secondary fracture in the shale formation, whose permeability is orders of magnitude larger than that of the shale-rock matrix and thus has a nearly constant pressure \cite{ref2}. The center-to-center width $d$ of the nanopore is 2 nm since pores with such small size are abundant in many shale formations \cite{ref2, ref30, ref31}. Because of the finite size of the methane molecules and the wall atoms, the accessible pore width in MD model $W_p$ is about 1.62 nm. The pore length $L$ is 202.5 nm. The left side of the gas bath is bounded by a piston plate, whose atoms move only in the $x$-direction. To maintain a constant pressure $P_{bath}$ inside the gas bath, an external force $F_{ext}=P_{bath} L_y L_z$ is applied on the piston plate ($L_y$ and $L_z$ are the plateâ€™s width in $y$- and $z$-directions, respectively). The external force is distributed on the atoms of the piston plate such that the plate remains planar. Hereafter, the MD model shown in Fig. \ref{fig1} will be termed nanopore-piston model for brevity. 

The methane molecules are modeled as structureless, spherical molecules. The interactions between a methane molecule â€œ$f$â€ and any another atom $j$ in the system are modeled using the Lennard-Jones (LJ) potential
\begin{equation}
  \phi_{fj} = 4 \epsilon_{fj} \left[ \left(\frac{\sigma_{fj}}{r}\right)^{12}- \left(\frac{\sigma_{fj}}{r}\right)^6 \right]
  \label{eq2}
\end{equation}
where $\sigma_{fj}$ and $\epsilon_{fj}$ are the LJ parameters for the interaction pair, and $r$ is the distance between the two molecules. The piston plate is modeled as a square lattice of carbon atoms (lattice spacing: 0.3 nm). We confirmed that the simulation results are independent of the mass for piston plate atoms. The pore walls are modeled as semi-infinite slabs constructed from a FCC lattice oriented in the $\langle111\rangle$ direction (following Ref. \onlinecite{ref11}, the lattice constant is taken as 0.54 nm). Explicitly modeling all wall atoms will incur significant computational cost. To overcome this difficulty, only the innermost layer of the wall atoms (i.e., the layer in contact with methane molecules) is explicitly modeled. Since the discreteness of atoms behind the innermost layer is barely felt by the gas atoms inside the pore due to the short-ranged nature of the LJ potential, these wall atoms are treated collectively as an implicit slab, and any methane molecule inside the system interacts with this implicit slab of wall atoms (the shaded region in Fig. \ref{fig1}) via the LJ 9-3 potential
\begin{equation}
  \phi_{f-iw} = \frac{2}{3} \pi \rho_w \epsilon_{fw} \sigma_{fw}^3 \left[\frac{2}{15} \left(\frac{\sigma_{fw}}{r_0}\right)^9 - \left(\frac{\sigma_{fw}}{r_0}\right)^3 \right]
  \label{eq3}
\end{equation}
where $\rho_w$ is the number density of the wall atoms in the implicit slab, $\epsilon_{fw}$ and $\sigma_{fw}$ are the LJ parameters for interactions between wall atoms and methane molecules, and $r_o$ is the distance between the methane particles and the surface of the implicit slab. Although such a wall model does not reflect the rich chemical nature of shale formation, it can well capture the notable adsorption of gas on pore walls. The LJ parameters for the methane molecules and the wall atoms are taken from Refs. \onlinecite{ref34} and \onlinecite{ref35}, which describe gas-silica interactions. While this choice is less relevant to surfaces of organic matters, it is reasonable for mineral surfaces, which are commonly found in shale gas formations. More importantly, the gas-wall potential parameters adopted here lead to a gas adsorption similar to those found in prior studies of gas adsorption in shale formations \cite{ref36}. This allows us to delineate the essential physics of gas recovery in presence of gas adsorption on pore walls, which is one of the important aspects in gas recovery from shale formations. The LJ parameters used in this study are summarized in Table \ref{forcefields}. 

	\begin{table}[htpb]  
	\caption{Lennard-Jones parameters for interactions between molecules.}
	\begin{center} \begin{tabular}{l l l } \hline
	Pair             &$\sigma$ (nm)  & \ \ \ \ $\epsilon/k_B$ (K) \ \ \ \  \ \\ \hline
	methane-methane \ \ \ \   &0.3810 	                & \ \ \ \  148.1 \\ 
    methane-wall          &0.3355                   & \ \ \ \ 207.2 \\
	methane-wall (weak) \ \ \ \ \ \ \ \ \ \  \       &  0.3355                  & \ \ \ \ 119.6 \\ \hline
	\end{tabular} 
	\end{center} \label{forcefields}
	\end{table} 	

Simulations are performed using the Lammps code \cite{ref37} with a time step size of 2 fs. Cutoff lengths of 1.2 nm and 1.5 nm are used in the calculation of the methane-methane interactions and the methane-wall interactions, respectively. In each simulation, the number of methane molecules inside the system is kept as constant and the pore walls are fixed in position. The dimensions of the simulation box are kept constant, but the volume of the gas bath increases during gas recovery simulation as its pressure is maintained by the piston wall (note that the simulation box is large enough in the x-direction that the piston plate never crosses its left boundary). The temperature of the methane molecules is maintained at 373 K using the Nose-Hoover thermostat. To explore the mechanisms of the gas recovery process, we designed several simulation cases with different pressures in the gas bath ($P_f=25,100,200$ bar) while the initial gas pressure in the nanopore is fixed at $P_o = 250$ bar. These choices of the initial gas pressure in the nanopore, the pressure in the gas bath, and the gas temperature, have been used in recent simulations \cite{ref2, ref30, ref31} and are relevant to the situations in some shale formations.

Each simulation consists of three steps. In step A, the blocker atoms at the pore entrance are removed and $F_{ext}$ on the piston plate is set to a value corresponding to the gas bath pressure $P_o$. The system is then evolved to equilibrium. In step B, the blocker atoms at the pore entrance are reinstated and $F_{ext}$ on the piston plate is set to a value corresponding to the gas bath pressure $P_f$. The system is then evolved till equilibrium is reached in the gas bath. In step C, the blocker atoms at the pore entrance are removed (this moment is defined as $t=0$) while $F_{ext}$ on the piston plate remains at the value set in the previous step. The system is evolved for 200ns to study the gas recovery from the nanopore. During each simulation, the density, pressure, and temperature of gas in both the nanopore and the gas bath are computed on-the-fly. Each simulation was repeated three times with different initial configurations to obtain reliable statistics. 

\section{Results and Discussion}
\label{sec:results}
\subsection{Qualitative and quantitative aspects of gas recovery}
We quantify the recovery of gas from the nanopore by computing a dimensionless recovery fraction ($RF$):
\begin{equation}
RF(\tilde t) = n/N	 
\label{rf_def}
\end{equation}
where $n$ is the cumulative production of mole of gas from the nanopore, and $N$ is the mole of gas in the nanopore at the initial state ($t = 0$). $\tilde t$ is a dimensionless time $\tilde t = t/t_c$, where $t_c$ is the characteristic time for the gas recovery process. Since the Knudsen diffusion often dominates the transport of gas in the narrow pore considered here and gas recovery from nanopores is often a diffusive process \cite{ref31}, $t_c$ is chosen as \cite{ref2}
\begin{equation}
t_c=  \frac{4 L^2}{D_m^o/K_n^0}
\label{tref_def}
\end{equation}
where $L$ is the pore length. $K_n^0$ and $D_m^o$ are the Knudsen number and the molecular diffusion coefficient of the gas molecules inside the pore at the initial state, respectively. $D_m^o$ is given by 
\begin{equation}
D_m^o= \frac{2}{3} \frac{\sqrt{m k_B T/ \pi}}{A \rho_0}
\label{dm0}
\end{equation}
where $m$ is the mass of a single gas molecule, $k_B$ is the Boltzmann constant, $T$ is the absolute temperature inside the pore, $A$ is the cross-sectional area of the gas molecule, and $\rho_0$ is the density of pore gas at the pressure and temperature corresponding to those found initially inside the pore. With the choice of pore width and gas in our system ($m=2.66\times10^{-26}$kg, $A \approx 0.45$ nm$^2$, $W_p =$ 1.62 nm, $P_0=250$ bar, $T=373$ K, $\rho_0=168$ kg$\cdot$m$^{-3}$), The initial mean free path length of gas inside the pore is $\lambda_0=0.26$ nm and the corresponding Knudsen number is $K_n^0=0.16$. Consequently, the gas transport inside the nanopore is not in the Knudsen regime initially. However, once the gas recovery operation starts, the gas density inside the pore drops and becomes non-uniform along the pore. The gas transport transitions into the Knudsen regime in some positions of the pore. In particular, at the pore entrance, the local pressure rapidly approaches the gas pressure in the bath and the Knudsen number increases. For example, if the gas bath pressure is $P_f=25$ bar, the Knudsen number of gas at the pore entrance reaches 1.12 soon after the gas recovery process starts.

\begin{figure}[hptb]
  \begin{center}
    \epsfig{file=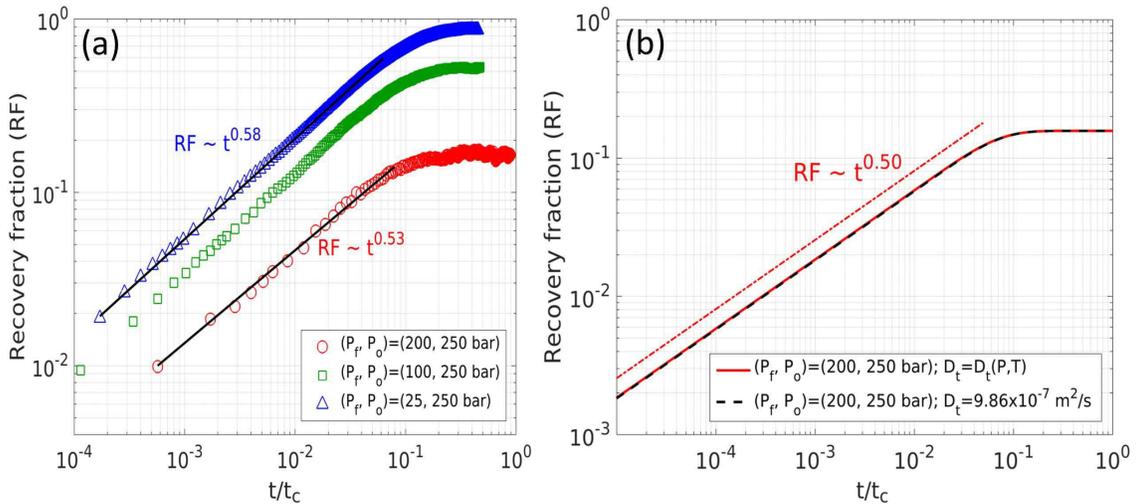,width=6.0in}  
    \caption{(a) The cumulative gas recovery computed by MD simulations for different gas bath pressure ($P_f =$ 25, 100, 200 bar) but the same initial gas pressure inside the pore ($P_o$ = 250 bar). (b) Cumulative recovery fraction calculated by continuum model with total diffusion coefficient taken as a constant or the pressure-dependent form given by Equation \eqref{Dt_def}.
    }
    \label{fig2}
  \end{center}
\end{figure}

Figure \ref{fig2}a shows the evolution of gas recovery fraction for several cases in which the initial pressure in the pore is fixed at $P_o$=250 bar while the gas bath pressure $P_f$ varies from 25 to 200 bar. The gas recovery fraction increases rapidly during the early stage of operation but increases very slowly for $\tilde t \gtrsim$ 0.1, similar to what was reported in prior works \cite{ref2, ref31}. Hereafter, we focus on the initial stage of gas recovery since shale gas recovery at the late stage is usually too slow to be practically useful. For $P_f=200$ bar, $RF(\tilde t) \sim \tilde t^{0.53}$ during the early stage of operation, indicating that the gas recovery is quite close to a diffusive process \cite{ref2, ref31}, in which $RF(\tilde t) \sim \tilde t^{0.50}$. However, for $P_f=$ 25 and 100 bar, $RF(\tilde t) \sim \tilde t^{0.58}$, indicating that the gas recovery deviates more greatly from the diffusive scaling. This super-diffusive scaling differs from the prediction by prior theories qualitatively, but it resembles some reported experimental data. Specifically, the average gas production rate obtained from 25 wells by Baihly \emph{et al}. \cite{ref32} showed that the production rate decays as $t^{-0.4}$ and the accumulative gas recovery fraction increases on $t^{0.6}$, which is faster than $t^{0.5}$. 

Since gas recovery is close to the diffusive scaling law predicted by previous theories for $P_f =$ 200 bar, we further examine whether the gas recovery under this condition can be predicted by prior theories \cite{ref2}. Following these theories, for the gas recovery from the slit pore shown in Fig. \ref{fig1}, the gas density averaged across the pore, $\rho$, is governed by a one-dimensional model
\begin{equation}
    \frac{\partial \rho}{\partial t} + \nabla \cdot \boldsymbol{j} = q
    \label{eq7}
\end{equation}
where $\rho$ is the gas density inside the pore. $q$ is a source term, which is equal to zero when the desorption of gas from pore walls is not considered. $\boldsymbol{j}$ is the gas flux given by $\boldsymbol{j} = -D_t d\rho/dx$, where $D_t$ is the total diffusion coefficient. Note that isothermal transport is assumed in Equation \eqref{eq7}. According to kinetic theories  \cite{ref2}, we have
\begin{equation}
D_t(P,T) = \frac{D_m^0}{Kn} \left( \frac{2}{7} \left( \frac{\rho}{\rho_0 Kn}+4\right) + \frac{1}{1+\rho/(\rho_0 Kn)} \right)
\label{Dt_def}
\end{equation}
where $\rho$ is the gas density. The first term in the right-hand side is the gas advection including slippage effect and the second term is molecular and Knudsen diffusion under linear diffusion conditions. For the gas recovery simulated in our MD simulations, the initial and boundary conditions of Equation \eqref{eq7} are given by
\begin{equation}
\begin{array}{r@{}l}
    \rho(x,0) = &\ \ \rho_0 \\ 
    \rho(0,t) = &\ \ \rho_f \\
    \frac{\partial \rho}{\partial t}(L,t) =&\ \  0 
\end{array}
\label{eq9}
\end{equation}
where $\rho_0$ is the initial gas density inside the nanopore, and $\rho_f$ is the final density inside the pore. Solving Equations \eqref{eq7}-\eqref{eq9} leads to a prediction of the gas recovery from the nanopore. During gas recovery operations, the gas density varies along the pore. It follows from Equation \eqref{Dt_def} that the total diffusion coefficient of the gas molecules inside the pore varies temporally and spatially during these operations. Nevertheless, as shown in Fig. \ref{fig2}b, the $RF(t)$ predicted by simulations based on Equation \eqref{Dt_def} agrees very well with simulations in which $D_t$ is taken to an appropriate constant (hereafter we denote this constant the effective diffusion coefficient $D_{eff}$; physically, we expect $D_t(P_f) < D_{eff} < D_t(P_0)$). When the total diffusion coefficient is taken as a constant $D_{eff}$, Equation \eqref{eq7} can be solved analytically \cite{ref38}, and the gas recovery at short time is given by
\begin{equation}
RF(t) = 2(1-\rho_f/\rho_0) \sqrt{\frac{D_{eff}}{\pi L^2}} t^{1/2}
\label{ana}
\end{equation}
Fitting the gas recovery data from MD simulations under the conditions of $(P_0,P_f)$ = (250 bar, 200 bar) to Equation \eqref{ana}, one extracts an effective diffusion coefficient $D_{eff} = 4.62\times 10^{-7}$ m$^2$/s, which is smaller than the total diffusion coefficient of gas given by Equation \eqref{Dt_def} at both $P_0=$ 250 bar ($10.20\times10^{-7}$ m$^2$/s) and $P_f = $200 bar (9.27$\times10^{-7}$ m$^2$/s). The smaller $D_{eff}$ extracted from MD data suggests that, for the situation examined here, the gas recovery rate is smaller than that predicted by the classical theories.

The above results indicate that, depending on the operating conditions, the gas recovery from the narrow pore considered here can exhibit qualitatively different behavior (e.g., super-diffusive gas recovery) or quantitatively lower gas recovery rate compared to those predicted by classical theories. These discrepancies can be attributed to the strong adsorption of gas molecules on the solid wall and the gas desorption during gas recovery process, which are often neglected in the classical theories.

\subsection{Importance of gas-wall interactions for gas transport in nanopores}

We first study why the gas recovery predicted by MD simulations is slower than that predicted by the classical theory; in another word, why the effective diffusion coefficient of gas molecules in the nanopore extracted from gas recovery data is smaller than the total diffusion coefficient given by Equation \eqref{Dt_def}. To this end, we compute the total diffusion coefficient of the gas molecules in nanopore using MD simulations and examine its dependence on the strength of the gas-wall interaction.

\begin{figure}[hptb]
  \begin{center}
    \epsfig{file=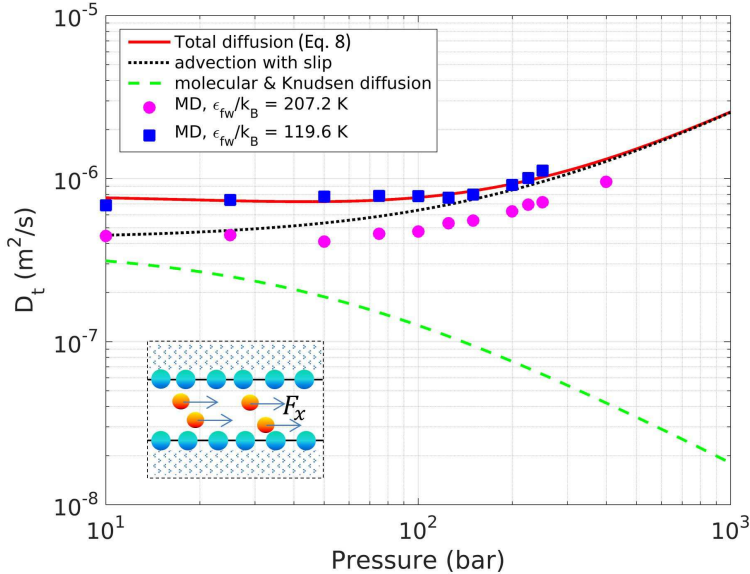,width=4.0in}  
    \caption{Comparison of the total diffusion coefficient of gas inside a 1.62 nm-wide nanopore determined using force-driven MD simulations and kinetic theory model (Equation. \eqref{Dt_def}). The inset shows a schematic of the MD system used for studying the gas transport in nanopores.
    }
    \label{fig3}
  \end{center}
\end{figure}

To compute the total diffusion coefficient of gas molecules inside the nanopore, we built a separate MD system consisting of a nanopore (it is periodical along the $x-$direction) and the gas molecules inside it (see inset of Fig. \ref{fig3}). The width of the pore, the structure of pore walls, and the average gas density inside pore are identical to those in the nanopore-piston system. A constant force of $F_x = 0.238$ pN is applied to each gas molecule in the $x-$direction, and the average velocity of gas molecules is computed. Since the total diffusion coefficient is defined based on the constitutive law of $\boldsymbol{j} = -D_t(\rho) \partial \rho/\partial x$, it cannot be computed directly from the above force-driven simulations. Instead, it is determined using the Darken equation \cite{ref39}
\begin{equation}
D_t = \frac{k_B T}{F_x} \langle v_x \rangle \left( \frac{\partial \ln f}{\partial \ln \rho}\right)_T = \frac{m}{F_x} \langle v_x \rangle \frac{\rho}{\rho_b} \left( \frac{\partial P}{\partial \rho}\right)_T
\label{darken}
\end{equation}
where $\langle v_x \rangle$ is the average velocity of gas molecules in the pore under a driving force of $F_x$ and $f$ is the fugacity. $\rho_b$ is the density in the gas bath, $\left(\frac{\partial P}{\partial \rho}\right)_T$ and $\frac{\rho}{\rho_b}$ are computed from the relation between the density and pressure of the gas confined inside the pore, which is determined during step A of the gas recovery simulation for the nanopore-piston system.

Figure \ref{fig3} shows the total diffusion coefficient  of the gas molecules confined in the pore when they interact with the pore walls with normal strength ($\epsilon_{fw}/k_B=$ 207.2 K) or with reduced strength ($\epsilon_{fw}/k_B=$ 119.6 K). For comparison, the predictions by Equation \eqref{Dt_def} are also shown. When pressure is less than $\sim$100 bar, molecule diffusion, Knudsen diffusion, and advection with slip all play an important role; for pressure larger than $\sim$100 bar, advection with slip dominates the overall diffusion coefficient. In agreement with the kinetic theory, $D_t$ increases with pressure, which can be attributed to the important role of advection in nanopores with high gas density. For gas-wall interactions with normal strength, $D_t$ increases from 6.30$\times$10$^{-7}$ to 7.17$\times$10$^{-7}$ m$^2$/s as the pressure inside the pore increases from 200 to 250 bar. We note that the $D_{eff}$ determined from the gas recovery simulation operating with $(P_0, P_f) =$ (250 bar, 200 bar) is 4.62$\times$10$^{-7}$m$^2$/s, which is still somewhat smaller than the above range. This can be understood that the simulation with $(P_0, P_f) =$  (250 bar, 200 bar) is not a strictly diffusive gas recovery, hence the effective diffusion coefficient extracted from fitting the gas recovery to Equation \eqref{ana} is not strictly accurate. Over the entire range of pressure investigated here, $D_t$ computed from MD simulation with normal gas-wall interactions is always smaller than that predicted by Equation \eqref{Dt_def}. When the gas-wall interaction is reduced, $D_{t}$ increases and approaches toward that predicted by the kinetic theories. This suggests that the strong gas-wall interactions cause the slower transport of gas inside the narrow pore compared to the kinetic theory predictions. Strong gas-wall interactions slow down the transport of the gas molecules confined in nanopores because they lead to adsorption of gas molecules on the pore walls. Under a given driving force, the transport of these adsorbed molecules is slower compared to the free gas in the center of the pore because their friction with pore wall atoms. Such an effect is especially obvious in narrow pores because, the fraction of the wall-adsorbed gas inside a pore increases as the pore size decreases. While these effects are not included in the diffusion model given by Equation \eqref{Dt_def}, which is derived from kinetic theories, they are explicitly included in MD simulations. These effects can also be captured well using the theoretical models developed previously \cite{ref29}. Since the diffusion coefficient can be computed very efficiently using these theoretical models, these models may be used to replace the diffusion model given by Equation \eqref{darken} in future shale gas researches.

\subsection{Importance of gas adsorption/desorption for gas recovery}

The strong gas-wall interactions lead to significant adsorption of gas molecules on the pore wall. During a gas recovery operation, the pressure inside the pore drops and those adsorbed gas molecules gradually desorb from the pore wall. Below we show that such gas desorption causes the super-diffusive gas recovery (i.e., $RF(t) \sim {t}^n$, with $n > 0.5$) observed in Fig. \ref{fig2}.

\begin{figure}[hptb]
  \begin{center}
    \epsfig{file=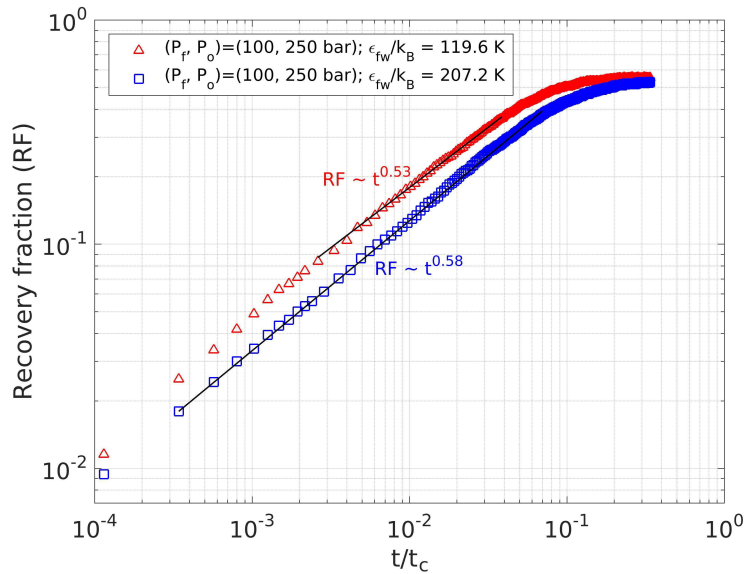,width=4.0in}  
    \caption{MD prediction of the evolution of gas recovery fraction for
      normal and weak gas-wall interactions (see Table \ref{forcefields}).
      The gas bath pressure and initial pressure inside pore are the same, i.e., $(P_0,P_f)=$ (250 bar, 100 bar).
    }
    \label{fig4}
  \end{center}
\end{figure}

The importance of gas desorption in the super-diffusive gas recovery is consistent with the observation that, for a fixed $P_0=250$ bar in the nanopore, super-diffusive gas recovery is observed only when the gas bath pressure is much lower than $P_0$, i.e., when $P_f=$ 25 or 100 bar but not when $P_f= $ 200 bar. This is because gas desorption from pore walls is minor during gas recovery if the initial pressure of gas inside the nanopore is close to the gas bath pressure. Along a similar line, if the adsorption of gas on pore wall is weak at $t = 0$, the subsequent desorption of gas during gas recovery will be weak, and the super-diffusive gas recovery should be suppressed. To test this argument, we perform gas recovery simulations in the nanopore-piston system with reduced gas-wall interactions ($\epsilon_{fw}/k_B=$119.6 K), and results are shown in Fig. \ref{fig4}. We observe that for $(P_0,P_f)=$ (250 bar, 100 bar), $RF(\tilde t) \sim \tilde t^{0.53}$, which is close to the $RF \sim \tilde t^{0.5}$, except at very short time ($\tilde t < 2.5\times 10^{-3}$). Indeed, since adsorption/desorption cannot be completely eliminated even if the reservoir pressure is close to the initial pore pressure or when gas-wall interactions are weak, weak deviation from the diffusive behavior is expected. While gas recovery operating under the same $(P_0,P_f)$ but with normal gas-wall interaction strength exhibits more distinct super-diffusive scaling.

To further ascertain that gas desorption causes the super-diffusive gas recovery shown in Fig. \ref{fig2}, we incorporate this effect into the classical gas recovery model and examine the gas recovery computed using the improved model. Specifically, we adopt the method reported by Shabro \emph{et al}. \cite{ref33} by setting the source term in Equation \eqref{eq7} to be
\begin{equation}
q = \chi (J_d - J_a)
\label{eq12}
\end{equation}
where $\chi$ is the surface to volume ratio of the slit pore, and $J_d$ and $J_a$ are the gas desorption and adsorption fluxes, respectively. $J_d$ is given by
\begin{equation}
J_d = n(P,T) k_d
\label{eq13}
\end{equation}
where $n(P,T)$ is the number density of gas adsorbed on pore walls. $k_d = k_0 \exp(-E_d/k_BT)$ is the gas desorption rate ($k_0$ is a pre-factor and $E_d$ is the desorption energy). The adsorption flux $J_a$ is given by:
\begin{equation}
J_a = (n_{\infty} - n(P,T)) k_a P
\label{eq14}
\end{equation}
where $k_a$ is the adsorption rate and $n_{\infty}$ is the number density of the gas adsorption sites on the pore wall.

\begin{figure}[hptb]
  \begin{center}
    \epsfig{file=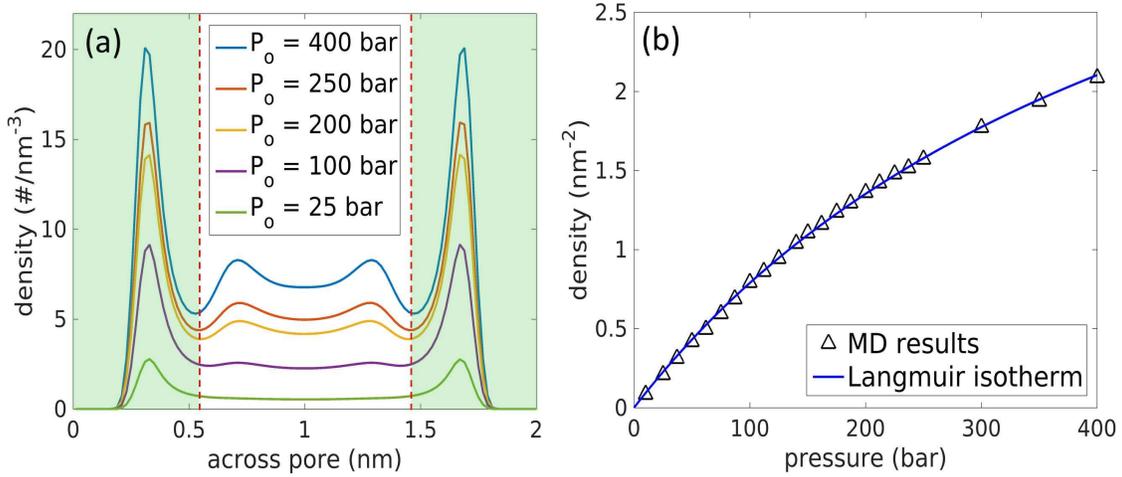,width=6.0in}  
    \caption{(a) Gas density profile across the nanopore computed using MD simulations under different pore pressure. Gas within the shaded regions is considered as adsorbed gas. (b) The gas adsorption isotherm determined using the gas density data in (a) and its fitting to the Langmuir isotherm.
    }
    \label{fig5}
  \end{center}
\end{figure}

The gas recovery model described by Equations \eqref{eq7}, \eqref{eq9} and \eqref{eq12}-\eqref{eq14} is hereafter referred to as the adsorption-desorption-transport (ADT) model. The model input parameters include $D_{eff}$, $k_0$, $E_d$, $n_{\infty}$, $k_a$, and $k_d$. The total gas diffusion coefficient $D_{eff}$ is taken as a constant because the gas recovery from nanopore can be predicted quite well when the spatial and temporal variation of gas diffusion coefficient inside the nanopore during gas recovery is neglected, providing a suitable value of $D_{eff}$ is used (cf. Fig. \ref{fig2}b). 
$D_{eff}$ and $k_0$ are taken as adjustable parameters in the ADT model. The desorption energy $E_d$ is determined by computing the depth of the potential well for the gas molecules in the layer next to the wall, and it is found to be 4.36 kJ/mol for the walls in our system. Of the other parameters, $n_{\infty}$ and $k_a/k_d$ can be obtained by studying the thermodynamics of gas adsorption on the pore wall.
Specifically, we first divide the gas inside the pore into â€œfree gasâ€ and â€œadsorbed gasâ€ and compute the isotherm of the adsorbed gas inside the pore. As shown in Fig. \ref{fig5}a, the peaks in the shaded region correspond to the adsorption of gas molecules on the surface. This adsorption is driven by the attractive van der Waals forces between gas molecules and the wall: the potential energy of a gas molecule close to the wall is lower than in the pore center due to its van der Waals interactions with the wall, hence it is energetically favorable for gas molecules to adsorb on the wall. 
In order to determine the gas adsorption on the pore wall, we partition the pore space into two â€œwallâ€ zones ($z <$ 5.9 \AA \  and $z >$ 14.1 \AA) and an â€œinteriorâ€ zone (5.9 \AA \ $< z <$ 14.1 \AA). 
The adsorbed gas density $n$ is computed using 
$n = \int_{0}^{5.9 \AA} {\rho} (z) dz$ for various pore pressure considered. We next fit the computed adsorption-pressure relation to the Langmuir isotherm \cite{ref40}
\begin{equation}
n(P,T) = a \frac{b P}{1+ b P}
\label{langmuir}
\end{equation}
where $a$ and $b$ are constants. Using Equations \eqref{eq13} and \eqref{eq14}, one readily shows that $a = n_{\infty}$ and $b = k_a / k_d$. Figure \ref{fig5}b shows that the gas adsorption on the pore walls can be described very well by the Langmuir isotherm, with $n_{\infty} = 4.73$ nm$^{-2}$ and $k_a/k_d = 0.02$ MPa$^{-1}$.
 
\begin{figure}[hptb]
  \begin{center}
    \epsfig{file=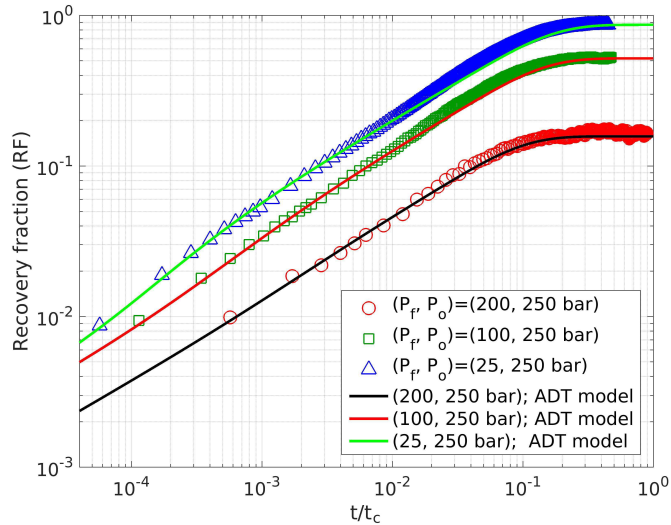,width=3.5in}  
    \caption{Comparison of the gas recovery fraction predicted by MD simulations and the adsorption-desorption- transport model for $(P_0,P_f)=$ (250, 25 bar), $(P_0,P_f)=$ (250, 100 bar), $(P_0,P_f)=$ (250, 200 bar).
    }
    \label{fig6}
  \end{center}
\end{figure}

With the above parameters, we solved the ADT model to predict the gas recovery from nanopore studied in our MD simulations with $(P_0,P_f)=$ (250, 25 bar), and compare the result with that computed using MD simulations. Figure \ref{fig6} shows that, with $k_0 =$ 0.84$\times10^{13}$ s$^{-1}$ and $D_{eff}^{25-250} =$ 4.06$\times10^{-7}$ m/s$^2$, the gas recovery predicted by the continuum model agrees quite well with our MD results. We note that the desorption pre-factor $k_0$ used is within the range used in previous studies \cite{ref41}. In addition, the effective diffusion coefficient used above close to the range of $D_t$ computed through separate MD simulations with normal gas-wall interaction strength ($\epsilon_{fw}/k_B = $207.2 K) ($D_t =$ 4.10 $\sim$ 7.17$\times10^{-7}$ m/s$^2$ from 25 to 250 bar, see Fig. \ref{fig3}). The good agreement between the ADT model and the MD simulations confirms that the significant gas desorption impacts near the wall surface is responsible for the super-diffusive gas recovery observed in MD simulations. In addition, while the ADT model described above has been used in previous studies \cite{ref28, ref33} for shale gas recovery, it has not been validated quantitatively using more fundamental methods (e.g., MD simulations) to our knowledge. Given that the ADT model divides the gas within a nanopore into free gas and adsorbed gas (with no lateral mobility but finite exchange with free gas) while even adsorbed gas molecules have finite lateral mobility, it is not clear \emph{a priori} that the ADT model can predict the gas recovery accurately. This is especially true in narrow pores in which the division between the free gas and adsorbed gas is less sharp compared to wide pores. The results shown in Fig. \ref{fig6} thus provides support for the effectiveness of the ADT model despite its rather drastic simplification of the real gas adsorption, desorption, and transport processes in nanopores.

\begin{figure}[hptb]
  \begin{center}
    \epsfig{file=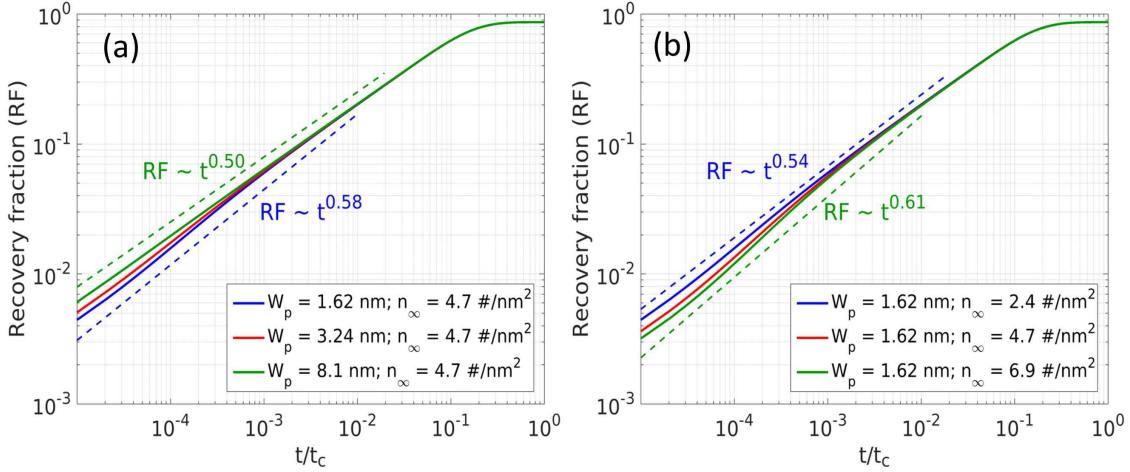,width=6.0in}  
    \caption{Effects of pore width $W_p$ (a) and maximal gas adsorption capacity of pore walls nâˆž (b) on the gas recovery from nanopores predicted using the adsorption-desorption transport model for $(P_0,P_f)=$ (250 bar, 25 bar).
    }
    \label{fig7}
  \end{center}
\end{figure}

Because the ADT model can capture the gas recovery behavior from nanopores well, we next use it to examine how the scaling law of gas recovery is affected by the properties of the nanopores. In particular, we vary the pore width $W_p$ and the maximal adsorption capacity of the pore wall $n_{\infty}$ to study how they affect the deviation of gas recovery from the classical diffusive behavior. Figure \ref{fig7}a shows that, for a fixed $n_{\infty} = 4.73$ nm$^{-2}$, the gas recovery becomes more â€œdiffusiveâ€ as the pore width increases and the classical diffusive scaling is recovered in pores with width of 8.1 nm. Figure \ref{fig7}b shows that, for a fixed $W_p =$ 1.62 nm, gas recovery still exhibits notable super-diffusive behavior even when the maximal gas adsorption capacity reduces to $n_{\infty} = 2.4$ nm$^{-2}$. Together, these results suggest that the super-diffusive gas recovery found in this work will likely occur in shale formations with pore size up to $\sim$ 5 nm.

\section{Conclusions}
\label{sec:conclusions}
 We perform MD simulation to investigate the gas extraction from single nanopores. The cumulative gas recovery and the effective diffusion coefficients for gas transport inside pores are studied. The results showed that, in very narrow pores, the strong gas-wall interactions can change the gas recovery behavior both quantitatively and qualitatively. These interactions slow down the gas recovery rate because they cause the total diffusion coefficients of the gas inside nanopores to be smaller than those predicted by kinetic theories. Additionally, these interactions lead to significant adsorption of gas molecules on the pore walls and cause a super-diffusive behavior $RF \sim \tilde t^{0.58}$. We show that, even in very narrow pores, the super-diffusive gas recovery behavior can be captured quantitatively using the coupled adsorption-desorption-transport model despite its rather drastic simplification of the physical processes during gas recovery. Parametric studies using the coupled adsorption-desorption-transport suggest that, at the single-pore level, super-diffusive gas recovery occurs in slit pore with width up to $\sim$5 nm if the difference between the initial pressure inside nanopore and the pressure in the large gas bath (or equivalently large fractures) is large. These results demonstrate that gas recovery can exhibit super-diffusive behavior at the single-nanopore level. While this behavior is observed in a highly idealized pore model, it should hold for nanopores with different shape and surface chemistry. This is because the super-diffusive scaling observed here originates from the coupled gas adsorption/desorption with gas transport inside nanopores, which is a rather generic feature of most pores in shale formations. Whether the super-diffusive gas recovery at single nanopore level is the decisive factor leading to the super-diffusive gas recovery behavior observed in shale wells, however, is not clear at present. A definitive answer of this question requires simulations that take into account factors such as the polydispersity of pore sizes, the diverse chemical nature of nanopore walls, and the connectivity between pores in realistic shale formations.

\section*{Acknowledgments}
The authors thank the ARC at Virginia Tech for generous allocation of computing time on the 
NewRiver and BlueRidge clusters. 

%

\end{document}